\documentclass[manuscript,screen]{acmart}

\AtBeginDocument{%
  \providecommand\BibTeX{{%
    \normalfont B\kern-0.5em{\scshape i\kern-0.25em b}\kern-0.8em\TeX}}}

\setcopyright{acmcopyright}
\copyrightyear{2026}
\acmYear{2026}

\usepackage{graphicx}

\usepackage{amsmath}

\newcommand{\remark} {\noindent {\sc \underline{Remark:} }}

\newcommand{\ip}     {\noindent {\bf Input: }}
\newcommand{\op}     {\noindent {\bf Output: }}
\newcommand{\proc}     {\noindent {\bf Processors: }}

\newcommand{\ps}[3]{\mbox{\sc Prefix$(#1,#2,#3)$}}
\newcommand{\mr}[2]{\mbox{\sc Merge$(#1,#2)$}}

\newcommand{\sr}[2]{\mbox{\sc Search$(#1,#2)$}}
\newcommand{\st}[2]{\mbox{\sc Sort$(#1,#2)$}}

\newcommand{\pcmssort}[4]{\mbox{{\sc PCM\_SSORT}$_{#1}(#2,#3,#4)$}}
\newcommand{\segr}{\mbox{{\sc Slice\_and\_Stratify}}}
\newcommand{\stratify}{\mbox{{\sc Stratify}}}
\newcommand{\ssort}[3]{\mbox{{\sc CREW\_SSORT}$_{#1}(#2,#3)$}}

\newcommand{\msort}[3]{\mbox{{\sc PEM\_SORT}$_{#1}(#2,#3)$}}

\newtheorem{theorem}{\bf Theorem}

\newtheorem{lemma}{\bf Lemma}
\newtheorem{cor}{\bf Corollary}

\begin{document}

\title[Sorting in PRAM, PCM and PEM]{Sorting in PRAM, Parallel Comparison Model and Parallel External Memory Model}

\author{Gopalan Sajith}
\authornote{This author contributed the most to this research.}
\email{sajith@iitg.ac.in}
\orcid{0000-0002-5412-5525}
\author{Manas Kumar Panda}
\authornote{Both of these authors contributed equally to this research.}
\email{manas.panda@iitg.ac.in}
\author{Ranajit Senko}
\authornotemark[2]
\email{ranajit@iitg.ac.in}
\affiliation{%
  \institution{Indian Institute of Technology Guwahati}
  \streetaddress{Department of Computer Science and Engineering}
  \city{Guwahati}
  \state{Assam}
  \country{India}
  \postcode{781039}
}

\renewcommand{\shortauthors}{G. Sajith, M. K. Panda and R. Senko}

\begin{abstract}
  We study the problem of sorting on the Parallel Random Access Machine (PRAM), Parallel Comparison Model (PCM) and Parallel External Memory Model (PEM). On PRAM, we present a Concurrent Read Exclusive Write algorithm that runs in $O(\log n)$ time with $n$ processors; our algorithm is very different from Cole's Merge Sort. On PCM, we present the first optimal algorithm for sorting after AKS; our algorithm runs in $O(\log n/\log\alpha)$ comparison rounds with $n\alpha$ processors, for $1\leq\alpha\leq n$. On PEM, we present a Concurrent Read Exclusive Write algorithm that sorts $N$ elements in $O\left(\log\frac{N}{B}\right)$ I/O rounds with $p\geq \frac{N}{B}$ processors. 
\end{abstract}

\ccsdesc[500]{Theory of computation~Shared memory algorithms}
\ccsdesc[500]{Theory of computation~Sorting and searching}

\keywords{external memory algorithms, models of computation}


\maketitle

\section{Introduction}

Sequential algorithms, studied in traditional algorithmics, are designed to run on a single processor \cite{ahu, clrs}. 
Improvements in processor speeds have hitherto enabled us to solve increasingly larger problem instances using such algorithms. 
However, processor speeds are approaching their inherent physical limits, making it difficult to achieve further speedups through 
       sequential algorithms alone. 
As technological advances continue to spawn ever larger problem instances, the use of multiple processors will become essential. 
Algorithms specifically designed to run on multiple processors are known as parallel algorithms.

\subsection{Models of Parallel Computation}

In this paper, we study some parallel algorithms designed for the following models of parallel computation: the Parallel Random Access Machine 	(PRAM) \cite{j-ipa-92}, the Parallel Comparison Model (PCM) \cite{j-ipa-92, AzarVishkin1987}, and the Parallel External Memory model (PEM) \cite{sitchinava-esa10, Arge2, Arge1}. 

The PRAM model envisions a set of synchronous processors that collectively share a random-access memory; each of these processors resembles the processor of the Random Access Machine model \cite{ahu}. 
Each PRAM instruction comprises three phases: Read, Compute, and Write. 
Processors transition through these phases synchronously, ensuring that all processors are in the same phase at any given time. 
The shared memory is assumed to be large enough to accommodate all the necessary data.
Multiple processors can concurrently read from or write to the memory. 
Depending on how conflicts from concurrent reads or writes to the same memory location are handled, 
    different variants of the PRAM model exist: the EREW (exclusive read, exclusive write) PRAM prohibits multiple processors from reading or writing the same memory location simultaneously; the CREW (concurrent read, exclusive write) PRAM allows concurrent reads but disallows concurrent writes; and the COMMON CRCW (concurrent read, concurrent write) PRAM permits concurrent reads and concurrent writes, provided that all processors write the same value. 
Many other variants of CRCW PRAM can be found in the literature \cite{j-ipa-92, SajithS00}. 
The performance of a PRAM algorithm is typically expressed in terms of the number of processors used and the amount of time, measured in instruction cycles, required to complete the computation.

In the Parallel Comparison Model (PCM), each comparison incurs unit cost, while all other operations are considered free. A comparison round involves multiple comparisons performed concurrently. The performance of a PCM algorithm is expressed in terms of the number of comparison rounds. Lower bounds established for PCM are applicable to all PRAM models.

The Parallel External Memory (PEM) model \cite{sitchinava-esa10, Arge2} combines elements from both PRAM and external memory (EM) models \cite{Aggvitt}.
Like PRAM, PEM employs a set of synchronous processors, and, like EM, it assumes that the data exceed the capacity of the main memory.  
In PEM, processors share an external memory, while each processor has an individual main memory of size $M$.
Computation proceeds in synchronized read and write I/O rounds.
In each read (write) I/O round, a block of size $B$ is transferred from (resp., to) the shared external memory to (resp., from) the 
	main memory of each processor.
The performance of a PEM algorithm is typically measured in terms of the number of processors used and the amount of time, measured in
	I/O rounds, required to complete the computation.

\subsection{Sorting and Merging}

Sorting a set of items drawn from a linearly ordered set is the problem of arranging them in ascending or descending order. 
When comparisons are the only operations allowed on the items, sorting is referred to as comparison-based. 
The sequential time complexity of comparison-based sorting is $\Theta(n \log n)$ \cite{ahu,clrs}. 

To merge two sorted arrays is to combine them into a single one.
In the multiway merging problem, we combine multiple sorted arrays into one. 
When there are $k$ arrays, the problem is called $k$-way merging. 
The cases $k = 2$ and $k = n$ correspond to ordinary merging and sorting, respectively. 
When only comparisons are allowed on the input elements, these problems are also referred to as comparison-based.
Comparison-based 2-way merging requires only $O(n)$ time in the sequential setting \cite{ahu,clrs}.

Sorting, merging, and multiway merging are fundamental operations that are widely used 
      in databases, search engines, big data processing, computer graphics, and network log analysis, to name a few.
This makes them important and extensively studied in algorithmics.

\subsubsection{On PRAM and related models}

Batcher gave $O(\log^2 n)$-depth, $O(n \log^2 n)$-cost comparator networks for sorting~\cite{k-ss-73}. 
Later, Ajtai, Koml\'{o}s, and Szemer\'{e}di \cite{aks-scLnps-83,p-pct-87,p-cn-90} built a sorting network (known as AKS) that has 
    $O(\log n)$-depth, $O(n \log n)$-size, but with large constants hidden in the $O$-notation.
The same bounds apply on EREW PRAM. 
Cook et al. \cite{cdr-ultbpramwsw-86} showed that with any number of processors, OR needs $\Omega(\log n)$ time on CREW PRAM, implying the 
    same for sorting. 
Hence, AKS sorting is optimal even for CREW PRAM. 
Cole gave a simpler pipeline-based  EREW PRAM algorithm with $O(\log n)$ time, $O(n)$ processors, and small constants \cite{Cole2}. 

Azar and Vishkin proved that on the parallel comparison model, sorting with $p \geq n$ processors needs at least $\Omega\left(\frac{\log n}{\log(p/n)}\right)$ steps \cite{AzarVishkin1987}. Alon gave a matching upper bound (via AKS) with large constants \cite{AzarVishkin1987}. Cole’s merge sort achieves a time of $O\left(\frac{\log n}{\log(p/n)} \cdot \log\log(p/n)\right)$. Columnsort \cite{l-tbcps-85} runs in $O\left((\log n/\log p)^{3.419}\right)$ time, and Cubesort \cite{cs-c-92} in $O\left(2^{\log^*(n/p)} (\log n/\log p)^2\right)$ time with $2 \leq p \leq n$ processors.

In CRCW PRAM, Beame and Hastad \cite{bh-obdpCP-89} showed that parity needs $\Omega\left(\frac{\log n}{\log\log p}\right)$ time, implying that sorting needs \(\Omega\left(\frac{\log n}{\log(p/n)} + \frac{\log n}{\log\log p}\right)\) time with $p \geq n$ processors. No algorithm that matches this lower bound is known. The best known algorithm runs in
\(
O\left(\frac{\log n}{\log(p/n)} (\log\log(p/n))^5 \cdot 2^{O(\log^* n - \log^*(p/n)+1)} + \frac{\log n}{\log\log p} \right)
\)
time \cite{hr-fdaeps-93}, while Cole's merge sort on CRCW PRAM takes $O\left(\frac{\log n}{\log\log(p/n)}\right)$ time.

Sajith \cite{Sajith} and Sajith and Saxena \cite{SajithSsax} gave a parallel merge sort algorithm, that, unlike Cole's merge sort~\cite{Cole2} avoids pipelining and features small constant factors. In the parallel comparison model with $n\alpha$ processors ($2 \leq \alpha \leq n$), it sorts $n$ items in
\(
\frac{\log n}{\log \alpha} \cdot 2^{O(\log^{*}(\log n/\log \alpha))}
\) comparison rounds.  On a CREW PRAM with $n$ processors, this algorithm runs in $\log n \cdot 2^{O(\log^{*}n)}$ time, and on a CRCW PRAM with $n\alpha$ processors, in 
\(
\frac{\log n}{\log\log \alpha} \cdot 2^{O(\log^{*}(\log n/\log \alpha))}
\) time.

Comparison-based $2$-way merging takes $O(n)$ time sequentially.  
In parallel, Batcher gave $O(\log n)$ depth, $O(n\log n)$ size merging networks~\cite{k-ss-73}, which are optimal due to an $\Omega(n\log n)$ lower bound~\cite{k-ss-73}.  On an EREW PRAM, Hagerup and R\"{u}b~\cite{hr-omsEP-89} used Batcher's networks to get an $O(\log n)$ time optimal algorithm. 
A reduction from searching ~\cite{s-ps-85} shows $2$-way merging needs $\Omega(\log n)$ time on EREW PRAM.  
Borodin and Hopcroft~\cite{bh-rmspmc-85} showed merging needs $\Omega\left(\log\frac{\log n}{\log(p/n)}\right)$ time on the parallel comparison model with $p \geq n$ processors.  
This matches the CREW PRAM upper bound~\cite{bh-rmspmc-85,Valiant}.  
Kruskal~\cite{k-smspc-83} improved it to $O(\log\log n)$ time, optimally.

Sequential $k$-way merging needs $\Omega(n\log k)$ time~\cite{Wen}.  
Using heaps, it can be done in $O(n\log k)$ time~\cite{k-ss-73}.  
Lee and Batcher~\cite{lb-mmn-94} extended odd-even merge to $O(\log n \log k)$ depth for $k$-way merging.  
Wen~\cite{Wen} and Sajith~\cite{Sajith} gave $O(\log n)$ time optimal CREW PRAM algorithms.

\subsubsection{On PEM}

Arge et al.~\cite{Arge1} introduce the Parallel External Memory (PEM) model and establish a lower bound of 
 \(\Omega(\log (N/B))\)
parallel I/O rounds for sorting. Aggarwal and Vitter~\cite{Aggvitt} show a lower bound of \(\Omega\!\left(\frac{N}{B}\log_{M/B}\frac{N}{B}\right)\) in the external memory model, implying \(\Omega\!\left(\frac{N}{pB}\log_{M/B}\frac{N}{B}\right)\) in PEM. 

Thus, the combined lower bound is
\(
\Omega\!\left(\frac{N}{pB}\log_{M/B}\frac{N}{B} + \log (N/B)\right).
\)

They present distribution sort and merge sort algorithms that match this bound for many parameter values; the merge sort achieves an I/O complexity of $O \left( \frac{N}{pB} \log_{d} \frac{N}{B} \right)$,
where \(d = \max \left\{ 2, \min \left\{ \sqrt{\frac{N}{p}}, \frac{M}{B} \right\} \right\}\).

Cole and Ramachandran~\cite{Cole} propose SPMS, a deterministic, resource-oblivious sorting algorithm with \(O(N \log N)\) work, \(O \left( \frac{N}{B} \log_{M} N \right)\) cache complexity under tall-cache assumptions, and \(O(\log N \cdot \log \log N)\) parallel time. SPMS minimizes false sharing, runs efficiently under work-stealing schedulers, and scales well on modern multicore architectures without manual tuning.

Sitchinava and Zeh present a PEM merge algorithm that achieves an I/O complexity of \(O \left( \frac{N_1 + N_2}{pB} + \log p \right)\) using \(p\) processors~\cite{pbt12}.

\subsubsection{Other results on PRAM and PEM}

Algorithms on the PRAM model are well studied; see the textbook by J{\'a}J{\'a}~\cite{j-ipa-92}. 
In the PEM model, Arge et al.~\cite{Arge2} applied list ranking to graph problems using Delayed Pointer Processing, achieving \(\Theta(\text{sort}_P(N))\) parallel I/Os. This approach is central to solving connected components, minimum spanning tree, and biconnectivity problems in PEM. Jacob et al.~\cite{jacob} established that permuting lower bounds apply for list ranking in PEM and provided tight algorithms for a wider range of parameters. Sitchinava~\cite{Nodari1} investigated distribution sweeping and proposed a parallel buffer tree with I/O-optimal performance for range and membership queries. Subsequently, Sitchinava and Zeh~\cite{pbt12} proposed the Parallel Buffer Tree, a batched I/O-optimal data structure supporting insertions, deletions, membership and range queries with \(O(\frac{N}{pB}\log_{M/B}\frac{N}{B}+K/pB)\) parallel I/Os, where $K$ is the size of the output. Lieber~\cite{Lieber} proposed a PEM-efficient list ranking algorithm that matches known sorting bounds and introduced the Guided Interval Fusion (GIF) problem, establishing a stronger \(\Omega(\log^2 N)\) lower bound in specific PEM settings.

\subsection{Our Results}

In this paper, we present efficient parallel algorithms for the sorting problem under three different computational models. Our main results are the following:

\begin{enumerate}
    \item an $O(\log n)$ time algorithm for sorting using $n$ processors on a CREW PRAM.
    
    \item an $O \left( \frac{\log n}{\log \alpha} \right)$ time algorithm for sorting using $n \alpha$ processors on the Parallel Comparison Model (PCM), where $2 \leq \alpha \leq n$.
    
    \item an algorithm for sorting in the Parallel External Memory (PEM) model with $p\geq \frac{N}{B}$ processors that requires $O \left( \log \frac{N}{B} \right)$ I/O rounds under the CREW constraint.
\end{enumerate}

\subsection{Organization of this paper}

The remainder of the paper is organized as follows. Section~\ref{sec:prelim} presents the necessary preliminaries and formal definitions used throughout the paper. Section~\ref{sec:stratify} introduces the key procedure \textit{Stratify}, which forms the core of the proposed algorithms. In Section~\ref{sec:crewpram}, we describe an efficient algorithm for sorting on the CREW PRAM model. Section~\ref{sec:pcm} develops a sorting algorithm for the Parallel Comparison Model (PCM), focusing on minimizing the number of parallel comparison rounds. Section~\ref{sec:crewpem} extends the approach to the CREW PEM model and presents parallel algorithms for both sorting and multiway merging in an I/O-efficient manner. Finally, Section~\ref{sec:conclusion} concludes the paper with a summary of contributions and potential directions for future work.

\section{Preliminaries}
\label{sec:prelim}

Some notation and preliminary concepts are presented in this section.

The following definitions are from Cole \cite{Cole2}.
For three keys $x,y$ and $z$ drawn from a linearly ordered set,  we say that $y$ is {\em between}
        $x$ and $z$, 
	if $x \leq y < z$.
For a key $x$ and a sorted array $L$, the {\em rank\/} of $x$ in $L$ is the
	number of elements in $L$ that are less than or equal to $x$.
For two sorted arrays $L_{1}$  and $L_{2}$, $L_{1}$ is said to be {\em ranked\/} in $L_{2}$
	if, for every element $x$ of $L_{1}$, we know the rank of $x$ in $L_{2}$;  
	$L_{1}$ and $L_{2}$
	are said to be {\em cross ranked\/} if they are ranked into each other.

The rest of this section draws on Sajith \cite{Sajith} and Sajith and Saxena \cite{SajithSsax}.

We use the following expressions: 
\begin{itemize}
	\item $\mr{n}{p}$: the time complexity of merging two sorted arrays of a total size of $n$ with $p$ processors
        \item $\ps{n}{p}{b}$: the time complexity of finding the prefix sums over an array of size $n$ of $b$-bit numbers using $p$ processors
	\item $\st{n}{p}$: the time complexity of sorting an array of size $n$ using $p$ processors
        \item $\sr{n}{p}$: the time complexity of searching for an item in a sorted array of $n$ items using $p$ processors
\end{itemize}
The model of computation on which the time complexity is measured will become clear from the context. 

We assume that for all the computational models considered, the upper bounds for merging, prefix summing, and sorting follow a
        ''regularity'' condition.
Specifically, larger problem sizes with the same processor-to-input ratio (i.e., $p/n$) should not complete faster.
That is, if $n_{1}\leq n_{2}$, then $\mr{n_{1}}{n_{1}\beta}\leq \mr{n_{2}}{n_{2}\beta}$; and the same principle applies to 
	sorting and prefix summing.

Over the course of this paper, we shall often encounter 
	the following  allocation problem $\mathcal R$ \cite{Sajith}. Given an array
	$A$ of $k$ elements $n_{0},\ldots,n_{k-1}$, where $\sum_{r=0}^{k-1}n_r=n$, 
        fill an array $P[0\ldots n-1]$ so that, for $0\leq i<n$, $P[i]=j$ iff $j$ is the smallest index 
        so that $\eta_j = \sum_{r=0}^{j}n_r > i$ (or, in other words, $\eta_{j-1} \leq i < \eta_j$).

Using prefix sum and merge, problem $\mathcal R$ can be
	solved as follows. 
Compute the array $E$ of the prefix sums of $A$. 
Let $D[i]=i$, for $0\leq i< n$.
Cross rank $D$ and $E$. Now, $P$ can be filled as required 
	in the problem: set $P[i]$ as the rank of $D[i]$ in $E$.
Hence, the following lemma of \cite{Sajith, SajithSsax}: 

\begin{lemma}
\label{lem:aloc}
The problem $\mathcal R$ can be solved in
	$O\left(\frac{n}{p} + \ps{k}{p}{\log n} + \mr{n}{p}\right)$ time when $p$ processors are available on the PRAM models and PCM.
\end{lemma}

Given arrays $X_0, \ldots, X_{k-1}$ of sizes $n_{0},\ldots,n_{k-1}$, respectively, where $\sum_{i=0}^{k-1} n_{i}=n$,
	and $p$ processors, on the PRAM models and PCM, we can assign one processor for every $\Theta(n/p)$ elements of the input using Lemma 1.
Imagine that the arrays $X_0, \ldots, X_{k-1}$ are sequenced in that order contiguously. 
For $1\leq l\leq p$, let the $l$-th processor handle the elements numbered $a_l=(l-1)\lceil\frac{n}{p}\rceil$ to
	$b_l=l\lceil\frac{n}{p}\rceil-1$ in this imagined sequence.
An algorithm for $\mathcal R$ can be used to translate $a_l$ and $b_l$ into pointers to the corresponding elements in the input.
Processor $l$ now knows the identity of all the elements it is to handle \cite{Sajith, SajithSsax}.

\section{Procedure Stratify}
\label{sec:stratify}

Our algorithms use a procedure called \stratify, which we describe at a high level in this section.
The details of its implementation on specific models will be discussed in other sections.
This is adapted from a similar procedure by Sajith and Saxena \cite{Sajith, SajithSsax}.

\vspace{0.2in}
\hrule
\vspace{0.1in}

\noindent {\bf Procedure $\stratify(n, k, X_0, \ldots, X_{k-1}, \sigma, \lambda)$}

\ip 
Parameters $n$ and $k$ are positive integers.
Array $X_{i}$ is of size $n_{i}$, for $0\leq i< k$, 
	and is sorted in increasing order;
	$\sum_{i=0}^{k-1} n_{i}=n$. 
Without loss of generality,
	we assume that the elements of the arrays are all distinct, because otherwise, for each $i$,
	the $j$-th element $x_{i,j}$ of $X_{i}$, can be replaced by a triplet 
	$\langle x_{i,j},i,j\rangle$,
	with the linear ordering extended lexicographically.
Parameters $\lambda$ and $\sigma$ are integers so that $\sigma< k\lambda < n$.

\op 
A permutation of the input elements, given in an array 
	$X[0\ldots n-1]$ so that $X$ has $s\in \left[ \frac{n}{\lambda}, \frac{2n}{\lambda}\right)$ (not necessarily sorted) 
	subarrays $Y_{j}(X)$ of size $|Y_{j}(X)| \in [\sigma, k\lambda+\sigma]$,  for $0\leq j< s$, given one after another contiguously,
	so that for all $j$ and all $j'<j$, each element of $Y_j(X)$ is larger than every element of 
	$Y_{j'}(X)$. 

\begin{description}

\item[\bf Step 1:]
Divide $X_{i}$, for $0\leq i< k$, into segments of size $\lambda$ each.
There are $s_{i}=\lceil \frac{n_{i}}{\lambda}\rceil$ segments in $X_{i}$.
If $n_{i}$ is not divisible by $\lambda$, the last segment is of a size smaller than $\lambda$.  
So the total number $s=\sum_{i=0}^{k-1} s_{i}$ of segments formed satisfies 
	$\frac{n}{\lambda} \leq s\leq \frac{n}{\lambda}+k < \frac{2n}{\lambda}$.
In each segment, designate its least element as a {\em bristle}. 
Let an array $X'_{i}$, for $0\leq i<k$, contain the set of bristles from $X_{i}$
	arranged in increasing order.

For $0\leq i<j< k$, cross-rank the bristle arrays
	$X'_{i}$ and $X'_{j}$.
Each bristle $b$ now has $k$ ranks, the rank $r_{i}(b)$ of $b$ in $X'_{i}$ for $0\leq i<k$,
	assigned to it. 

\item[\bf Step 2:]
For each bristle $b$, in parallel, add together all $r_{i}(b)$'s. 
Consequently, each bristle knows its rank among all the bristles;  that is, now, all the bristles 
	can be placed together in an array $L=\langle L_0, \ldots, L_{s-1}\rangle$ in ascending order.
Let $L_s=+\infty$.

\item[\bf Step 3:]
Each bristle $b$ now knows, for each $i$, the segment $\sigma_{i}(b)$ of $X_i$ into which it falls. 
For each $b\in L$ and for each array $X_{i}$, search $\sigma_{i}(b)$ in parallel to find the rank of $b$ in it.   
From this $b$ can infer its rank $R_{i}(b)$ in $X_{i}$. 

\item[\bf Step 4:]
For each bristle $b$,
	compute $I(b)=\sum_{i=0}^{k-1} R_{i}(b)$, the
	rank of $b$ in $X$, the output array.
\end{description}

\remark
For $0\leq j<s$, 
	there can be at most $\lambda$ elements in $X_{i}$, $0\leq i< k$,
	that are in the interval $[L_j, L_{j+1})$; these elements are deemed to be under the charge of $L_j$.
That is, $R_{i}(L_{j+1})-R_{i}(L_j)\leq \lambda$, for all $i$. 
Hence, $I(L_{j+1})-I(L_j)\leq k\lambda$.

\begin{description}
\item[\bf Step 5:]
If $\sigma>0$, then choose and remove (See Section~\ref{sec:stratpem} for details) some elements of $L$, and then compact the remaining elements of $L$ into an array $L'$ 
	(of size, say, $s'\leq s$) so that for $0\leq j<s'$, $\sigma\leq I(L'_{j+1})-I(L'_j)\leq k\lambda+\sigma$.
Set $L=L'$ and $s=s'$. Assume that $L_{s}=+\infty$.
\end{description}

\remark For each removed element, the elements that were under its charge are transferred to the largest smaller bristle that remains. (This step is relevant only for the PEM implementation.)

\begin{description}
\item[\bf Step 6:]
For $0\leq j<s$, and $0\leq i< k$, let 
	\[ \delta_{i}(L_j)=\left\{ \begin{array}{ll}
			   R_{i}(L_{j+1})-R_{i}(L_j) & \mbox{when $L_{j+1}\not \in X_{i}$}\\
			   R_{i}(L_{j+1})-R_{i}(L_j)-1 & \mbox{when $L_{j+1} \in X_{i}$}
                        \end{array}
	         \right.
	\]
Note that $\delta_{i}(L_j)$ is the number of elements that are larger than $L_j$ but smaller than
	$L_{j+1}$ in $X_{i}$.
    
Let $\langle\Delta_{0}(L_j),\ldots,\Delta_{k-1}(L_j)\rangle$ 
	be the prefix sums computed over
	$\langle\delta_{0}(L_j),\ldots,\delta_{k-1}(L_j)\rangle$; $\Delta_{-1}(L_j)$ is set to zero.

If $\delta_{i}(L_j)>0$, then copy $X_{i}[R_{i}(L_j) \ldots R_{i}(L_j)+\delta_{i}(L_j)-1]$ to
	$S_{L_j}[\Delta_{i-1}(L_j) \ldots \Delta_{i}(L_j)-1]$.
    
That is, we isolate all elements that are larger than $L_j$ but smaller than
	$L_{j+1}$ into an array $S_{L_j}$. 
\end{description}

\remark
Note that $\mid S_{L_j} \mid = I(L_{j+1})-I(L_j)-1 < k\lambda+\sigma$.

\begin{description}
\item[\bf Step 7:]
Copy $L_j$, for $0\leq j<s$, in parallel,
	into the $(I(L_j)-1)$-st location of $X$.
Copy $S_{L_j}$, for $0\leq j<s$, in parallel,
	into the subarray of $X$ of equal size that begins at index $I(L_j)$.
\end{description}

\remark
Array $X$ is now in the desired form; $L_j$ and $S_{L_j}$ together form 
	the required subarray $Y_{j}(X)$ of $X$. 
\vspace{0.2in}
\hrule
\vspace{0.1in}

\section{Sorting on a CREW PRAM}
\label{sec:crewpram}

We use Procedure Stratify to design an optimal CREW PRAM sorting algorithm, 
	which is described in this section.
We begin with an analysis of the procedure.   

\subsection{Procedure Stratify on a CREW PRAM when $\sigma=0$}
\label{sec:anastratify}


We can partition the input arrays and 
	form the bristle arrays in $O(1)$ time using $n$ processors, once
	they are assigned one per element. 
Assign one processor for every $\Theta(\frac{n}{p})$ elements of the input,
	using Lemma~\ref{lem:aloc}. 
Each processor can now sequentially select bristles from the elements assigned to it.
All this takes $O(\frac{n}{p}+\ps{k}{p}{\log n}+\mr{n}{p})$ time.	
Cross-ranking of the bristle arrays involves $\frac{k(k-1)}{2}$ merges, 
	each of size at most $s<\frac{2n}{\lambda}$, to	be performed in parallel.
Each bristle participates in $k-1$ merges.
Use Lemma~\ref{lem:aloc} to assign $\Theta(\frac{p}{s})$ processors per bristle;
	$1/k$-th of those processors will participate in each of the merges of the bristle. 
Solve each merge instance independently. 
The time taken is $O(\mr{\frac{n}{\lambda}}{\frac{n}{\lambda}*\frac{p}{sk}})=O(\mr{\frac{n}{\lambda}}{\frac{p}{k}})$.
 
Steps~2 and 4 each requires every bristle to perform an addition of size $k$,
	of $\log n$ bit numbers.  
Use Lemma~\ref{lem:aloc} to assign $\Theta(\frac{p}{s})$ processors per bristle. 
Find the sums using
	a prefix summation algorithm in $O(\ps{k}{\frac{p}{s}}{\log n})$ time. 
The total time taken is $O(\frac{n}{p}+\ps{k}{\frac{p}{s}}{\log n}+\mr{n}{p})$. 

Each bristle has to perform $k$ searches in Step~3, each over a segment of
	size $\lambda$; that is, a total of $sk$ searches have to be performed 
	in parallel. 
At most $\lfloor \frac{p}{sk} \rfloor$ processors are available per search.
Hence, we can finish this step in $O(\sr{\lambda}{\frac{p}{sk}})$ time.

Since $\sigma=0$, Step~5 takes $O(1)$ time. 

The subtractions in Step~6 can be performed in $O(1)$ time with $sk
	$ processors.
With $p$ processors, this would take 
	$O(sk/p)$ time.
Step~6 also requires every bristle to perform a prefix summation of size $k$,
	of $\log \lambda$ bit numbers. 
With a processor allocation similar to that in Step~2,
	the sums can be found in $O(\ps{k}{\frac{p}{s}}{\log \lambda})$ time. 
Since the bristles are now merged into array $L$, the allocation incurs
	only $O(1)$ time.

In Step~6, we also prepare the subarrays $S_b$, $b\in L$.
Each bristle $b$, for $0\leq i\leq k-1$, 
	if $\delta_{i}(b)\not =0$,
	tags element  $X_{i}[R_{i}(b)]$ with
	the address of the location $S_{b}[\Delta_{i-1}(b)]$; 
	note that $X_{i}[R_{i}(b)]$, the leftmost element of $X_i$ in $b$'s charge,
    is to be copied to 
	$S_{b}[\Delta_{i-1}(b)]$.
Assign $\Theta(\frac{p}{s})$ processors to each segment;
	and for each element find 
	the largest tagged element smaller than it in its array.
As the second element in the segment is guaranteed 
	to be tagged, the first being a bristle, this does not require any
	inter-segment information access;
	a prefix minima over each segment, done in parallel, will do. 
Thus, each element can compute the address of the cell it must be copied to.
So, the total time taken is
	$O(\ps{\lambda}{\frac{p}{s}}{\log\lambda})$. 

In Step~7, we copy the bristles and $S_b$'s into $X$ to form the $Y_j(X)$'s.
Since every element knows where it is to be copied, this takes $O(\frac{n}{p})$ time. 

On a PRAM that allows concurrent reads, 
	$\ps{n}{p}{b}=O(\frac{n}{p} + \log n)$ and $\mr{n}{p}=O(\frac{n}{p} + \log\log n)$ \cite{Valiant,j-ipa-92}. 
Hence, for a constant $c$,
	\[\ps{n^{c+1}}{pn^{c}}{b}=\Theta(\ps{n}{p}{b})\] 
	and
	\[\mr{n^{c+1}}{pn^{c}}=\Theta(\mr{n}{p})\]
That is, if $\lambda=k^{O(1)}$, then $\ps{\lambda}{\frac{p}{s}}{\log\lambda})=O(\ps{k}{\frac{pk}{s\lambda}}{\log n})$.

Also, as $s=\Theta(n/\lambda)$, if $\lambda\geq k$, then  
$\mr{\frac{n}{\lambda}}{\frac{p}{k}}=O(\mr{n}{\frac{p\lambda}{k}})=O(\mr{n}{p})$. 


Thus, we have the following lemma:

\begin{lemma}
\label{lem:stratify}
The procedure call `` $\stratify(n, k, X_0, \ldots, X_{k-1}, 0, \lambda)$'' with $k\leq\lambda=k^{O(1)}$ takes 
	\[O\left(\frac{n}{p}+\ps{k}{\frac{pk}{s\lambda}}{\log n} + \mr{n}{p} + \sr{\lambda}{\frac{p}{sk}}\right)\] time 
	with $p$ processors on a PRAM that allows concurrent reads.
\end{lemma}

\subsection{A Special Case of Stratify}
\label{sec:segre}

In this subsection we describe Procedure \segr. 
This procedure call takes as input an array with a small number of 
	large sorted subarrays given one after another contiguously,
	and produces as output a permutation of the input elements
	that has a large number of small subarrays so that
	all elements of a subarray are larger than every element of a 
	preceding subarray. 
The procedure is described below. 

\vspace{0.05in}
\hrule
\vspace{0.1in}

\noindent {\bf Procedure \segr$(n,k,A,\sigma,\lambda)$}

\vspace{0.1in}

\ip
$A$ is an array of $n$ distinct elements with $k$ 
	sorted subarrays $A_{i}$, $0\leq i< k$, 
	of size $n/k$ each, 
	given one after another contiguously, where $n$ is a multiple of $k\lambda$.

\op 	
The output of the procedure is a permutation of the input elements, given in array 
	$A$ itself, that would have $n/\lambda$ (not necessarily sorted) 
	subarrays $G_{j}(A)$ of size $g_{j}(A)$, $g_{j}(A) \leq k\lambda$,
	$0\leq j< n/\lambda$, given one after another contiguously,
	so that all elements of a subarray are larger than every element of a 
	preceding subarray.

\proc
The number of processors available is $p=n\beta$, $\beta>1$.

\begin{description}
\item[Step:] Invoke $\stratify(n, k, A_0, \ldots, A_{k-1}, \sigma, \lambda)$. Overwrite $A$ with the output.
\end{description}
\hrule
\vspace{0.1in}

\begin{lemma}
\label{lem:slstratify}
The procedure call `` $\segr(n, k, A, 0, k)$'' uses $n\beta\geq n$ processors and takes 
	\[O\left(\frac{1}{\beta}+\ps{k}{k\beta}{\log n} + \mr{n}{n\beta} + \sr{k}{\beta}\right)
	=O(\log k+\log\log n)\]
	time on a PRAM that allows concurrent reads.
\end{lemma}

\subsection{A CREW PRAM algorithm for sorting}
\label{sec:crew}

We construct a sequence of sorting algorithms 
	using the following idea:
\begin{quote}
	If there exists an $O(c \log n \log^{(h)}n)$ time $O(n)$ processors algorithm
	$\mathcal A$ for sorting $n$ items, where $h\geq 2$, for some $d$ such that $\frac{d}{c}=O(1)$,
 	there exists an $O(d\log n\log^{(h+1)}n)$ time $O(n)$ processors algorithm 
	$\mathcal B$ for sorting $n$ items. 
\end{quote}
A high level description of the new algorithm $\mathcal B$ 
	would be as follows:
\begin{quote}
   {\bf Choose a parameter $k$.}

   {\bf Divide the input array of size $n$ into $\frac{n}{k^2}$ subarrays of size $k^2$ each.}
    
   {\bf Sort each subarray using algorithm $\mathcal A$.}

{\bf for $j\leftarrow 1$ to $\log_k n-2$ do} {\em /* There are $\frac{n}{k^{j+1}}$ sorted subarrays of size $k^{j+1}$ each. */}
\begin{quote}

{\bf Make groups of $k$ of the existing subarrays.}

{\bf Apply procedure \segr\ on all groups in parallel.}
\begin{quote}
{\em /* The elements of each group now appear in a sequence of (not necessarily sorted) 
	subarrays of size at most $k^2$ each, so that all elements of a 
	subarray are larger than every element of a preceding subarray within the group. */}
\end{quote}

\noindent 
{\bf Sort the resultant subarrays in parallel using algorithm $\mathcal A$.} 
\begin{quote}
    {\em /* Now each group has merged into a single sorted array.
That is, now there are $\frac{n}{k^{j+2}}$ sorted subarrays of size $k^{j+2}$ each. */}
\end{quote}

\end{quote}

{\bf endfor}

{\em /* Now a single sorted array is left. */}
\end{quote}
For an appropriate choice of $k$, this scheme gives the desired 
	time bound.

We now give the construction scheme more formally.
The basic building block of the scheme is Valiant's algorithm that
	merges two arrays of a total size $n$ in 
	$O(\log\log n)$ time with $n$ processors on a CREW PRAM 
	\cite{Valiant,j-ipa-92}; 
	let $a$ be the constant hidden by the $O$-notation here.
Guided by a complete binary tree on $n$ leaves, proceeding bottom up, and using
	Valiant's algorithm at each level, 
	the standard merge sort thus runs in $a\log n\log\log n$ time 
	on an $n$ processor CREW PRAM \cite{j-ipa-92}.
This is the $0$-th algorithm in our sequence.
We denote this algorithm by \ssort{0}{I}{n}. 

For $h>0$, \ssort{h}{I}{n} is given below.

We partition \(I\) hierarchically as follows.
For level \(j \ge 0\), define the $i$-th subarray of size $k_h^{j+2}$ in $I$ by
\[ I^{(j)}_i(h) \;=\;
I\big[\, i k_h^{\,j+2} \ldots (i+1)k_h^{\,j+2}-1 \,\big],
\]
where \(0 \le j \le \log_{k_h}n - 2\) and \(0 \le i < n/k_h^{\,j+2}\).

Each subarray at level \(j\) is composed of \(k_h\) consecutive subarrays at level \(j-1\):
\[
I^{(j)}_i(h)
\;=\;
\bigcup_{t=ik_h}^{(i+1)k_h-1} I^{(j-1)}_t(h).
\]


\begin{figure}[h]
\label{ssfig}
\vspace{0.1in}
\hrule
\vspace{0.1in}

\begin{tabbing}
aaaa \= aaaa \= aaaa \= aaaa \= aaaa \= aaaa \= \kill
{\bf Procedure \ssort{h}{I}{n}}\\
begin \\
     \> $k_h \leftarrow (\sqrt{\log n})^{f(h,n)}$, where $f(h,n)=1/\log^{(\lfloor h/2\rfloor+3)}n$; \\
     \> for $0\leq i< n/k_h^{2}$ in parallel do~~~~ 
		/* $I^{(0)}_{i}(h)=I[ik_h^{2},\ldots,(i+1)k_h^{2}-1]$ */ \\
     \>      \> \ssort{h-1}{I^{(0)}_{i}(h)}{k_h^{2}}; \\
     \> for $j\leftarrow 1$ to $(\log n/\log k_h)-2$ do \\
     \> begin \\
     \>	     \> for $0\leq i< n/k_h^{j+2}$ in parallel do \\
     \>      \> begin \\
     \>      \>      \> \segr$(k_h^{j+2},k_h,I^{(j)}_{i}(h),0, k_h)$; \\
     \>      \>      \> for $1\leq x\leq k_h^{j+1}$ in parallel do \\
     \>      \>      \>   \> \ssort{h-1}{G_{x}(I^{(j)}_{i}(h))}{g_{x}(I^{(j)}_{i}(h))}; \\
     \>      \> end \\
     \> end \\
end 
\end{tabbing}

\vspace{0.1in}
\hrule
\vspace{0.1in}
\end{figure}
On a CREW PRAM, $\ps{n}{n}{b}$ is $O(\log n)$,
	$\mr{n}{n}$ is $O(\log\log n)$ \cite{j-ipa-92}, and
	$\sr{k}{1}$ is $O(\log k)$.
Thus, by Lemma~\ref{lem:slstratify}, with one processor per input element,
	the procedure call ``\segr$(k_h^{j+2},k_h,I^{(j)}_{i}(h),0,k_h)$'' takes 
	$O(\log k_h + \log\log k_h^{j+2})$ time on a CREW PRAM.
Let $S_{h}(n)$ be the time taken by \ssort{h}{I}{n} with 
	$n$ processors. 
Then, for $h>0$,    
	\[S_{h}(n) \leq  
		O(1) + S_{h-1}\left(k_h^{2}\right) +
		\left(\frac{\log n}{\log k_h}-2\right) 
		\left(O(\log k_h) +  O(\log\log n) + S_{h-1}\left(k_h^{2}\right)\right)\]
Since on a CREW PRAM,
	$\Omega(\log k)$ is a lower bound on sorting 
	$k^2$ elements  with any number of processors \cite{cdr-ultbpramwsw-86},
	$\log k = O\left(S_{h-1}\left(k^{2}\right)\right)$.
Thus, for sufficiently large constants $b$ and $c$,
\[S_{h}(n) \leq  
		b \cdot \frac{\log n}{\log k_h} \cdot \log\log n \ \ + \ \ 
		c \cdot \left(\frac{\log n}{\log k_h}\right) \cdot S_{h-1}\left(k_h^{2}\right)\]
In particular,
\begin{eqnarray*}
S_{1}(n)  & \leq &  
		b \cdot \frac{\log n}{\log k_1} \cdot \log\log n \ \ + \ \ 
		c \cdot \left(\frac{\log n}{\log k_1}\right) \cdot S_{0}\left(k_1^{2}\right) \\
          & \leq & 
		b \cdot \frac{\log n}{\log k_1} 
		  \cdot \log\log n \ \ + \ \ ca \cdot \frac{\log n}{\log k_1} \cdot \log k_1^{2} \cdot \log\log k_1^{2}
\end{eqnarray*}
Since $\log k_1 = \frac{\log\log n}{2\log\log\log n}$, this means that 
$S_{1}(n) \leq	2(b+ac) \cdot \log n \cdot \log^{(3)}n$.  

Similarly, 
\begin{eqnarray*}
S_{2}(n) & \leq & 
		b \cdot \frac{\log n}{\log k_2} \cdot \log\log n \ \ + \ \ 
		c \cdot \left(\frac{\log n}{\log k_2}\right) \cdot S_{1}\left(k_2^{2}\right) \\
	   & \leq & 
		b \cdot \frac{\log n}{\log k_2} \cdot \log\log n \ \ + \ \ 
	    2c(b+ac) \cdot \frac{\log n}{\log k_2} 
			\cdot \log k_2^{2}\log^{(3)}k_2^{2} 
\end{eqnarray*}
Since $\log k_2 = \frac{\log\log n}{2\log^{(4)}n}$, this means that 
$S_{2}(n)  \leq (2b+4ac^2+4bc) \cdot \log n\log^{(4)}n\leq a \cdot \log n\log^{(3)}n$, when
$(2b+4ac^2+4bc)/a \leq \frac{\log^{(3)}n}{\log^{(4)}n}$.

\begin{lemma}
\label{crewlemma}
For $h>2$, the procedure call \ssort{2(h-2)}{I}{n} sorts an array $I$ of size $n$,
	with $n$ processors, on a CREW PRAM, in
	$a\log n\log^{(h)}n$ time, when $(2b+4ac^2+4bc)/a \leq \frac{\log^{(h)}n}{\log^{(h+1)}n}$.
\end{lemma}

\proof
The proof is by induction. The basis is when $h=3$; this was argued above.
Suppose the procedure call \ssort{2(h-2)}{I}{n} sorts an array $I$ of size $n$,
	with $n$ processors, on a CREW PRAM, in
	$a\log n\log^{(h)}n$ time, when $(2b+4ac^2+4bc)/a \leq \frac{\log^{(h)}n}{\log^{(h+1)}n}$.
Arguing as in the basis case, we find that $S_{2(h-2)+1}(n)  \leq 4ab \cdot \log n\log^{(h+1)}n$, and
$S_{2(h-1)}(n)  \leq (2b+4ac^2+4bc) \cdot \log n\log^{(h+2)}n$, which is at most $a \cdot \log n\log^{(h+1)}n$, when
$(2b+4ac^2+4bc)/a \leq \frac{\log^{(h+1)}n}{\log^{(h+2)}n}$. \qed

Define $\nu_h(D)=\min\left\{ n : D \leq \frac{\log^{(h)}n}{\log^{(h+1)}n}\right\} $.

By Lemma~\ref{crewlemma}, when $n\geq \nu_h(D)$, for constant $D=(2b+4ac^2+4bc)/a$, \ssort{2(h-2)}{I}{n} sorts an array $I$ of size $n$,
	with $n$ processors, on a CREW PRAM, in
	$a\log n\log^{(h)}n$ time. 
In particular, when $\nu_3(D) \leq n < \nu_4(D)$, use \ssort{2}{I}{n} to sort;
    this runs in $a\log n\log^{(3)}n = d \cdot \log n$ time, for a constant $d$.
(Denote $\log^{(4)}n$ by $x$; then $D\leq \frac{x}{\log x}$ implies that $\log^{(4)}(\nu_4(D))=O(D\log D)=O(1)$, and hence $a\cdot\log^{(3)}(\nu_4(D))=O(1)$ as well.)
When $\nu_4(D) \leq n < \nu_5(D)$, use \ssort{4}{I}{n} to sort;
    this too runs in $a\log n\log^{(4)}n = d \cdot \log n$ time. 
(This is because $\log^{(4)} (\nu_5(D))=\log^{(3)} (\nu_4(D))=O(1)$.)
In general, when $\nu_{h}(D) \leq n < \nu_{h+1}(D)$, \ssort{2(h-2)}{I}{n} sorts 
    $n$ items in $d \cdot \log n$ time. 

Given a sorting instance of size $n$, from the above sequence of algorithms we choose the appropriate one to be invoked on it; the algorithms are structurally identical, and are parameterized by $k_h$. The parameters $k_h$ can be precomputed offline. For each fixed $n$, the resulting lookup table may be reused across instances. 

Thus we have the following theorem:

\begin{theorem}
Using one of the procedures \ssort{2h}{I}{n}, for $h\geq 1$,  an array $I$ of size $n$ can be sorted 
	with $n$ processors, on a CREW PRAM, in
	$O(\log n)$ time. 
\end{theorem}

\section{An Algorithm for sorting on the Parallel Comparison Model}
\label{sec:pcm}

A generalization of the CREW PRAM algorithm, 
	given in the previous section, works on the  
	 parallel comparison model.

Here again, the basic building block is Valiant's algorithm that
	merges two arrays of a size of $n$ each in 
	$O(\log\log_{\alpha}n)$ time with $p=n\alpha$ processors on the parallel comparison model \cite{Valiant,j-ipa-92} (for a processor advantage $p/n$ of $\alpha$): pick every $\sqrt{n}$-th element from the two arrays as bristles, cross-rank the bristles in $O(1)$ time, rank the bristles into the two original arrays in $O(1)$ time, and recurse with the subproblems created at the same processor advantage, with the recursion ending when the subproblems have sizes upperbounded by $\alpha$, are provided with at least a quadratic number of processors each, and therefore are solvable in $O(1)$ time.

Guided by a complete $\sqrt{\alpha}$-ary tree on $n$ leaves, proceeding bottom up, and using
	Valiant's algorithm at each level, 
	the standard merge sort runs in $O(\log_\alpha n\log\log_\alpha n)$ time 
	on an $n\alpha$ processor PCM.
(At each non-leaf in the recursion tree, a $\sqrt{\alpha}$-way merge has to be performed.
For example, at the root of the recursion tree a $\sqrt{\alpha}$-way merge of sorted arrays
	of size $n/\sqrt{\alpha}$ each is to be performed with $n\alpha$ processors.
This can be achieved through a pairwise cross-ranking of the arrays; there are $\Theta(\alpha)$
	pairs; $\Theta(n)$ processors can be assigned to each pair; the cross-ranking can be 
	achieved in $O(\log\log_{\alpha}n)$ time; as addition comes for free on PCM, the merge
	is achieved at no further cost.)
Let $a$ be the constant hidden by the $O$-notation here.
This is the first algorithm in our sequence.
We denote this algorithm by \pcmssort{0}{I}{n}{\alpha}. 

For $h>0$, \pcmssort{h}{I}{n}{\alpha} is given below.
We use the notation introduced in the last section. 
For level \(j \ge 0\), define the $i$-th subarray of size $k_h^{j+2}$ in $I$ by
\( I^{(j)}_i(h) \;=\;
I\big[\, i k_h^{\,j+2} \ldots (i+1)k_h^{\,j+2}-1 \,\big],
\)
where \(0 \le j \le \log_{k_h}n - 2\) and \(0 \le i < n/k_h^{\,j+2}\).
Each subarray at level \(j\) is composed of \(k_h\) consecutive subarrays at level \(j-1\):
\(
I^{(j)}_i(h)
\;=\;
\bigcup_{t=ik_h}^{(i+1)k_h-1} I^{(j-1)}_t(h).
\)

On the parallel comparison model, prefix sums come for free,
	$\mr{n}{n\alpha}=O\left(\log\left(\frac{\log n}{\log\alpha}\right)\right)$, and
	$\sr{n}{p}=O\left(\frac{\log n}{\log p}\right)$.
Thus, by Lemma~\ref{lem:slstratify}, with $\alpha$ processors per input element,
	the procedure call ``\segr$(k_h^{j+2},k_h,I^{(j)}_{i}(h),0,k_h)$'' takes 
	$O\left(\log\left(\frac{\log n}{\log\alpha}\right) + \frac{\log k}{\log\alpha}\right)$ time on a PCM.

\begin{figure}
\label{pcmssfig}
\vspace{0.1in}
\hrule
\vspace{0.1in}

\noindent

\begin{tabbing}
aaaa \= aaaa \= aaaa \= aaaa \= aaaa \= aaaa \= \kill
{\bf Procedure \pcmssort{h}{I}{n}{\alpha}}\\
begin \\
     \> $k_h \leftarrow \sqrt{\alpha}^{f(h,n,\alpha)}$, where
          $f(h,n,\alpha)=\frac{\log \log_\alpha n}{\log^{(\lfloor h/2\rfloor+2)}\log_\alpha n}$; \\ \\
     \> for $1\leq i\leq n/k_h^{2}$ in parallel 
		do /* $I^{(0)}_{i}(h)=I[(i-1)k_h^{2}+1,\ldots,ik_h^{2}]$ */ \\
     \>      \> \pcmssort{h-1}{I^{(0)}_{i}(h)}{k_h^{2}}{\alpha}; \\
     \> for $j\leftarrow 1$ to $(\log n/\log k_h)-2$ do \\
     \> begin \\
     \>	     \> for $1\leq i\leq n/k_h^{j+2}$ in parallel do \\
     \>      \> begin \\
     \>      \>      \> \segr$(k_h^{j+2},k_h,I^{(j)}_{i}(h),0,k_h)$; \\
     \>      \>      \> for $1\leq x\leq k_h^{j+1}$ in parallel do \\
     \>      \>      \>   \> \pcmssort{h-1}{G_{x}(I^{(j)}_{i}(h))}{g_{x}(I^{(j)}_{i}(h))}{\alpha}; \\
     \>      \> end \\
     \> end \\
end 
\end{tabbing}

\vspace{0.1in}
\hrule
\vspace{0.1in}
\end{figure}
Let $S_{h}(n,\alpha)$ be the time taken by \pcmssort{h}{I}{n}{\alpha} with 
	$n\alpha$ processors.
Then, for $h>0$,  
	\[S_{h}(n,\alpha) \leq  
		O(1) + S_{h-1}(k_h^{2},\alpha) +
		\left(\frac{\log n}{\log k_h}-2\right) 
		(O(\log_\alpha k_h) + O(\log\log_\alpha n) + S_{h-1}(k_h^{2},\alpha))\]
Since on a PCM, 
	$\Omega(\log_\alpha k)$ is a lower bound on sorting 
	$k^2$ elements with $k^2\alpha$ processors \cite{cdr-ultbpramwsw-86},
	$\log_\alpha k = O(S_{h-1}(k^{2},\alpha))$.
Thus, for sufficiently large constants $b$ and $c$,
\[S_{h}(n, \alpha) \leq  
		b \cdot \frac{\log n}{\log k_h} \cdot \log \log_{\alpha} n \ \ + \ \ 
		c \cdot \left(\frac{\log n}{\log k_h}\right) \cdot S_{h-1}\left(k_h^{2}, \alpha\right)\]

In particular,
\begin{eqnarray*}
S_{1}(n, \alpha)  & \leq &  
		b \cdot \frac{\log n}{\log k_1} \cdot \log\log_{\alpha} n \ \ + \ \ 
		c \cdot \left(\frac{\log n}{\log k_1}\right) \cdot S_{0}\left(k_1^{2}, \alpha \right) \\
          & \leq & 
		b \cdot \frac{\log n}{\log k_1} 
		  \cdot \log\log_{\alpha} n \ \ + \ \ ca \cdot \frac{\log n}{\log k_1} \cdot \log_{\alpha} k_1^{2} \cdot \log\log_{\alpha} k_1^{2}
\end{eqnarray*}

Since $\log_\alpha k_1 = \frac{\log\log_\alpha n}{2\log\log\log_\alpha n}$,
this means that 
$S_{1}(n) \leq	2(b+ac) \cdot \log_{\alpha} n \cdot \log^{(2)} \log_{\alpha}n$.

Similarly, 
\begin{eqnarray*}
S_{2}(n, \alpha) & \leq & 
		b \cdot \frac{\log n}{\log k_2} \cdot \log\log_{\alpha} n \ \ + \ \ 
		c \cdot \left(\frac{\log n}{\log k_2}\right) \cdot S_{1}\left(k_2^{2}, \alpha\right) \\
	   & \leq & 
		b \cdot \frac{\log n}{\log k_2} \cdot \log\log_{\alpha} n \ \ + \ \ 
	    2c(b+ac) \cdot \frac{\log n}{\log k_2} 
			\cdot \log_{\alpha} k_2^{2}\cdot\log^{(2)}\log_{\alpha} k_2^{2} 
\end{eqnarray*}

Since $\log_{\alpha} k_2 = \frac{\log\log_{\alpha} n}{2\log^{(3)}\log_{\alpha}n}$, this means that 
\[S_{2}(n,\alpha)  \leq (2b+4ac^2+4bc) \cdot \log_{\alpha} n\cdot\log^{(3)}\log_{\alpha}n
\leq a \cdot \log_{\alpha} n\cdot\log^{(2)}\log_{\alpha}n\], when
$(2b+4ac^2+4bc)/a \leq \frac{\log^{(2)}\log_{\alpha}n}{\log^{(3)}\log_{\alpha}n}$.

\begin{lemma}
\label{pcmlemma}
For $h>2$, the procedure call \pcmssort{2(h-2)}{I}{n}{\alpha} sorts an array $I$ of size $n$,
	with $n\alpha$ processors, on PCM, in
	$a\log_{\alpha}n\log^{(h-1)}\log_{\alpha}n$ time, when $(2b+4ac^2+4bc)/a \leq \frac{\log^{(h-1)}\log_{\alpha}n}{\log^{(h)}\log_{\alpha}n}$.
\end{lemma}

\proof
The proof is by induction. The basis is when $h=3$; this was argued above.
Suppose the procedure call \pcmssort{2(h-2)}{I}{n}{\alpha} sorts an array $I$ of size $n$,
	with $n\alpha$ processors, on PCM, in
	$a\log_{\alpha} n\cdot\log^{(h-1)}\log_{\alpha}n$ time, when $(2b+4ac^2+4bc)/a \leq \frac{\log^{(h-1)}\log_{\alpha}n}{\log^{(h)}\log_{\alpha}n}$.
Arguing as in the basis case, we find that $S_{2(h-2)+1}(n)  \leq 4ab \cdot \log_{\alpha} n\log^{(h+1)}n$, and
$S_{2(h-1)}(n)  \leq (2b+4ac^2+4bc) \cdot \log_{\alpha} n\log^{(h+2)}n$, which is at most $a \cdot \log_{\alpha} n\log^{(h+1)}n$, when
$(2b+4ac^2+4bc)/a \leq \frac{\log^{(h)}\log_{\alpha}n}{\log^{(h+1)}\log_{\alpha}n}$. \qed

Define $\nu_h(D,\alpha)=\min\left\{ n : D \leq \frac{\log^{(h-1)}\log_{\alpha}n}{\log^{(h)}\log_{\alpha}n}\right\} $.

By Lemma~\ref{pcmlemma}, 
when $n\geq \nu_h(D,\alpha)$, for $D=(2b+4ac^2+4bc)/a$, \pcmssort{2(h-2)}{I}{n}{\alpha} sorts an array $I$ of size $n$,
	with $n$ processors, on a PCM, in
	$a\log_{\alpha}n\cdot\log^{(h-1)}\log_{\alpha}n$ time. Arguing as we did in the case of CREW PRAM in the previous section, we conclude that 
when $\nu_{h}(D,\alpha) \leq n < \nu_{h+1}(D,\alpha)$, \pcmssort{2(h-2)}{I}{n}{\alpha} sorts 
    $n$ items in $d \cdot \log_\alpha n$ time. 

Given a sorting instance of size $n$, from the above sequence of algorithms we choose the appropriate one to be invoked on it; the algorithms are structurally identical, and are parameterized by $k_h$. The parameters $k_h$ can be precomputed offline. For each fixed $n$, the resulting lookup table may be reused across instances. 

Thus we have the following theorem:

\begin{theorem}
Using one of the procedures \pcmssort{2h}{I}{n}{\alpha}, for $h\geq 1$, an array $I$ of size $n$ can be sorted 
	with $n\alpha$ processors, on a PCM, in
	$O(\log_{\alpha} n)$ time. 
\end{theorem}

\section{Sorting and Multiway-merging on CREW PEM}
\label{sec:crewpem}

A PEM model is said to be {\em self-simulating,} 
	if for all $N$ and $N'$, $N\geq N'\geq 1$, a PEM of the
	model with a size of $N'$ can simulate a single I/O round of another PEM of
	the model with a size of $N$ in $O(N/N')$ I/O rounds.
All Exclusive Write PEMs 
	are self-simulating. 

For example, the write phase of an $N$-processor CREW PEM can be simulated on an $N'$-processor CREW PEM
	by letting the $N'$ ``actual'' processors simulate the $N$ ``virtual'' processors in groups of
	size $N'$ each. 
Since no two processors write in the same memory block,   
	it is feasible to stagger the write into several I/O rounds.

The cost or work of an algorithm that runs in $T$ I/O rounds using $p$
	processors is the time-processor product $pT$.
If $\mbox{Seq}(N)$ is the worst-case running time (in I/O rounds) of  the  fastest 
	known sequential external memory algorithm for a problem of size $N$, an optimal 
	PEM algorithm  for the same problem runs in  
	$O(\mbox{Seq}(N)/p)$ time (in I/O rounds) using $p$ processors.

The degree of parallelism (DoP) of an I/O round is the 
	number of I/Os in it. 

For a PEM algorithm presented in $T$ I/O rounds, where the DoP of the $i$-th I/O round is $w_{i}$,
	the total number of blocks transferred is $W=\Sigma_{i=1}^{T} w_{i}$.
If we use $P=\max (w_{i})$ processors, the algorithm runs in $T$ I/O rounds, but $PT$ may be $\omega(W)$.
If our model is self-simulating, and we have $p\leq P$ processors, then
	the $i$-th I/O round can be simulated in $\lceil w_{i}/p \rceil$ I/O rounds.
So, the total time (in I/O rounds) taken by the simulation is
	\[ T_{s} = \Sigma_{i=1}^{T} \lceil w_{i}/p \rceil
	      \leq \Sigma_{i=1}^{T} (w_{i}/p + 1) 
	      =     T + W/p
 	\]
With $p= \lceil W/T \rceil$, $T_{s} = O(T)$, and the total cost of the execution is $O(W)$ I/Os. 

Hence, as with PRAM algorithms, we can describe our PEM algorithms in a number of I/O rounds
	of varying DoPs, without bothering
	about the actual number of processors available,
	provided, our presentation makes clear 
	the DoP of each step.
We shall refer to this principle as the {\em Brent's Theorem}
	or {\em WT Scheduling Principle} as applied to PEM.

\subsection{Some Basic PEM Algorithms}
\label{basicpem}
Now we discuss PEM solutions for a number of algorithmic problems. 
These algorithms follow easily from the corresponding PRAM algorithms. 
For each problem, we consider two scenarios,
	that we call sparse and compact. 
In a sparse instance, we assume that each element of the input 
	(as well as the output) occupies the last word of a distinct block, with the rest of the block unused.
In a compact instance, the input as well as the output is presented in contiguous locations 
	of the shared secondary memory.
The compact presentation of an array of size $N$ occupies $N/B$ blocks. 

\subsubsection*{Prefix Sums}
The input is an array $A[0\ldots N-1]$ of elements, and the output is to be an array $D[0\ldots N-1]$ 
	such that $D[i]=\Sigma_{j=0}^{i} A[j]$, for $0\leq i\leq N-1$. 

If the problem instance is sparse (with element $A[i]$ occupying the last word of the $i$-th disk block, with the remaining $B-1$ words of that block unused), we can use the EREW PRAM 
	balanced binary tree algorithm on an EREW PEM without any change, save that
	each instruction in the PRAM algorithm should correspond to an I/O in the PEM algorithm.
This algorithm, which we call PREFIX-SUMS-1, takes $O(\log N)$ I/O rounds for a total of $O(N)$ I/Os.
Thus by Brent's Theorem, with $p\geq 1$ processors, this can be run in $O\left(\frac{N}{p}+\log N\right)$ I/O rounds.

Instead if the instance were compact, then we proceed in three stages.
In stage~1, for $0\leq i\leq N/B-1$, in parallel, find the local 
	prefix-sums of the $i$-th block of $A$.  
In stage~2, invoke PREFIX-SUMS-1 on the $N/B$ block sums.  
In stage~3, for $1\leq i\leq N/B-1$, in parallel, offset the local prefix sums of block $i$ (computed in stage~1) 
	with the global prefix sum of block $(i-1)$ (computed in stage~2) and write the values in $D$.
We call this above algorithm PREFIX-SUMS-2. 
It takes $O(\log \frac{N}{B})$ I/O rounds for a total of $O(N/B)$ I/Os.
Thus by Brent's Theorem, with $p$ processors, PREFIX-SUMS-2 can be
	run in $O\left(\frac{N}{pB}+\log\frac{N}{B}\right)$ I/O rounds.

This bound improves on the $O\left(\frac{N}{pB}+\log p\right)$ bound in \cite{Arge1} because 
	when $\frac{N}{pB}=\Omega(\log\frac{N}{B})$, the two bounds are identical, and
	when $\frac{N}{pB}=o(\log\frac{N}{B})$, $\log\frac{N}{B}=O(\log p)$.

\subsubsection*{Searching}
If the input array $A[0\ldots N-1]$ has a sparse presentation, 
	to find the rank $R$ of element $x$ in $A$, on a CREW PEM with $N-1$ processors, initialize $R$ to $0$, 
	and then for $0\leq i\leq N-2$, in parallel, if $(A[i] \leq x$ and $A[i+1]>x)$, set $R$ to $i+1$.
This algorithm, which we call SEARCH-1, runs in $O(N/p+1)$ I/O rounds with $p$ processors on a CREW PEM.

Again, if $A$ has a sparse presentation, with the occupied blocks spaced evenly in the shared external memory,
	on a CREW PEM with $p$ processors, we would use 
	a straightforward generalization of the ordinary binary search.
If $(N\leq p+1)$, then return SEARCH-1$(A,x)$.
Otherwise, divide $A$ into $p+1$ segments of size $\lceil \frac{N}{p+1} \rceil$ each; 
	in parallel, compare $x$ with the
	last element of each segment, thus identifying the segment in which $x$ falls; 
	and recurse with that segment.
The search space reduces by a factor of $(p+1)$ in each I/O round.
Therefore, this algorithm, which we call SEARCH-2, runs in 
	$O\left(\frac{\log N}{\log (p+1)}\right)$ I/O rounds.

If $A$ has a compact presentation, then first find the block that contains $x$ using SEARCH-2,
	and then search for $x$ within that block in one additional I/O. 
This algorithm, which we call SEARCH-3, runs in 
	$O\left(\frac{\log (N/B)}{\log (p+1)}\right)$ I/O rounds on a CREW PEM with $p$ processors.

\subsubsection*{Merging}

Suppose two sorted arrays $C$ and $D$ of sizes $N_1$ and $N_2$ are given in sparse
	presentation, with the occupied blocks spaced equally in the shared external memory. 
We can find, in parallel, for $0\leq i\leq N_1-1$, the rank of $C[i]$ in $D$, and 
	for $0\leq j\leq N_2-1$, the rank of $D[j]$ in $C$, using SEARCH-2, 
	with $p$ processors per item in $O(\log_{p+1} N)$ I/O rounds, where $N=N_1+N_2$.
We call this algorithm CROSS-RANK.

To merge $C$ and $D$ we can use Valiant's merge algorithm \cite{j-ipa-92}. 
Divide $C$ (resp., $D$) into segments each of size $\sqrt{N_1}$ (resp., $\sqrt{N_2}$).
Elect the first element from each segment as a bristle.
Cross rank the bristles in $O(1)$ I/O rounds using CROSS-RANK; this
	requires only $\sqrt{N_1}\sqrt{N_2}<N$ processors.
For each bristle in $C$ (resp. $D$), search the segment in $D$ (resp. $C$) in which it falls
	using $\sqrt{N_2}$ (resp. $\sqrt{N_1}$) processors.
Now the problem is partitioned.
Solve each partition (of a size at most $\sqrt{N_1}+\sqrt{N_2}$)
	recursively.
The number of I/O rounds taken here is $T(N) \leq T(c.\sqrt{N}) + d$ for some constants $c$ and $d$.
So, $T(N) = O(\log\log N)$.
The number of processors is linear.
We call this algorithm MERGE-1.

Using MERGE-1, an $O(N)$ cost $O(\log\log N)$ I/O rounds CREW PEM algorithm
	for merging can be designed \cite{j-ipa-92}.
Divide $C$ and $D$ into segments of size $\log\log N$ each. 
Elect the first element from each segment as a bristle.
Cross rank the bristles in $O(\log\log N)$ I/Os using MERGE-1; this
	requires only $N/\log\log N$ processors.
For each bristle in $C$ (resp. $D$), search the segment in $D$ (resp. $C$)
	using one processor.
Now the problem is partitioned.
Solve each partition (of a size at most $2\log\log N$) sequentially.
We call this algorithm MERGE-2.

Now we describe an algorithm we call MERGE-3, in which the input arrays are assumed to be in
	compact presentation. 
From each block of each array choose the first element as a {\em bristle}.
Say, $C'$ and $D'$ are the arrays of bristles from $C$ and $D$ respectively.
Arrays $C'$ and $D'$ are then cross-ranked using MERGE-2.
Now each bristle knows 
	the block on the other side in which it falls.
For each bristle search this block to find its rank on the other side.
We say, this is where a bristle ``reflects'' in the other array.
As in MERGE-2, entrust to each bristle $x$ the items consecutive in its own array
	from itself to either a bristle or 
	a reflection of a bristle whichever appears first,
	and the items consecutive in the other array
	from the reflection of $x$ to either a bristle or 
	a reflection of a bristle whichever appears first.
Each bristle is entrusted a maximum of $2B-1$ items.
Let each bristle merge its entrusted elements sequentially.

If a bristle $x$ is entrusted $\epsilon$ elements from its own array and $\delta$ items from the other array,
	then the smallest bristle larger than $x$ gets at least $B-\epsilon$ items, and the largest bristle
	smaller than $x$ gets at least $B-\delta$ items.
This means that in the target array $C$, a block can contain items entrusted to at most three bristles.
Once the bristles merge their entrusted elements in the main memory, therefore, writing them onto the external memory
	will take at most three I/O rounds with $N/B$ processors. 
So, with $p$ processors, MERGE-3 runs in $O(\frac{N}{pB}+\log\log \frac{N}{B})$ I/O rounds.

\subsubsection*{Compaction}

Suppose a sequence $s$ of $B$ elements  
	is given in an array $A$ of $l$ blocks in the shared external memory, allowing an arbitrary number of empty memory 
	locations between two consecutive elements.
Using a balanced binary tree algorithm $s$ can be compacted into a 
	single block in $O(\log l)$ I/O rounds for a total of $O(l)$ I/Os on an EREW PEM.
We call this algorithm COMPACT-1.

If instead of $B$, the sequence $s$ has a length of $N$, then the following algorithm can be used. 
\vspace{0.1in}
\hrule
\vspace{0.1in}

\noindent
{\bf COMPACT-2}
\vspace{0.1in}
\hrule
\vspace{0.1in}

\ip A sequence of elements $a_1, \ldots, a_N$ given in that order in an array $A[1\ldots l]$ of blocks in the shared external memory.
There could be an arbitrary number of empty memory locations between two consecutive elements.

\op An array $Z[1\ldots N/B]$ of blocks that contains $a_1, \ldots, a_N$ in that order.

\remark In the following, $C$, $D$, $E$, $F$, $G$, $J$ and $K$ are arrays, initialized to zero, in sparse presentation.

\begin{description}
\item[\bf Step 1:] For $1\leq i\leq l$, let $C[i]$ hold the number of elements in block $A[i]$.
Invoke PREFIX-SUMS-1 on $C$. Let $D$ hold the prefix-sums. Let $D[0]=0$.

\item[\bf Step 2:] pardo for $1\leq i\leq l$, if there exists a positive integer $k_i$ such that $D[i-1]<k_i B\leq D[i]$,
(as $D[i] - D[i-1] = C[i] \le B$, there exists at most one such $k_i$)
then set boolean $flag[i]$, move the first $k_i B-D[i-1]$ elements of $A[i]$ into $Aux[i]$, set $E[i]$ to $k_i+1$ and set $K[i]$ to 1.

\remark If an element $x$ of $A[i]$ is to be the last element of a block of $Z$, then the elements of $A[i]$ that precede $x$ along with $x$ are moved into $Aux[i]$ and deleted from $A[i]$. Now $A[i]$ is the first block with elements to be moved into the $(k_i+1)$-st block of $Z$. 

\item[\bf Step 3:] 
Compute the PREFIX-MAXIMA on $E$ into $J$.

\remark The elements remaining in $A[i]$ are to be moved to the $J[i]$-th block of $Z$.
For $i>1$, the elements in $Aux[i]$ are to be moved to the end of the $J[i-1]$-st block of $Z$.

\item[\bf Step 4:] Let $F[1]=1$;  pardo for $2\leq i\leq l$, $F[i]= flag(i)*i$. Compute the PREFIX-MAXIMA of $F$ into $G$.

\remark If $G[i]=j$, then $A[j]$ is the first block to contribute elements to the block of $A[i]$ in $Z$.

\item[\bf Step 5:] Set $K[1]=1$, and pardo for $2\leq i\leq l$, if $flag[i]=0$, then set $K[i]=i-G[i]+1$. 

\remark $A[i]$ is to be the $K[i]$-th block participating in the $J[i]$-th block of $Z$.

\item[\bf Step 6:] Parallely invoke $N/B$ instances of COMPACT-1; the output of the $i$-th instance will be written to $Z[i]$.

\item[\bf Step 7:] pardo for $1\leq i\leq l$, if $flag[i]$, then append $Aux[i]$ to the elements already present in block $Z[J[i-1]]$.
\end{description}
\vspace{0.1in}
\hrule
\vspace{0.1in}

Steps 2, 5 and 7 takes $O(1)$ I/O rounds, while the remaining take $O(\log l)$ I/O rounds. The cost of each I/O round is $O(l)$ I/Os.
Thus by Brent's Theorem, with $p$ processors, COMPACT-2 can be run in $O\left(\frac{l}{p}+\log l\right)$ I/O rounds.
If $l=O(N/B)$, then with $p$ processors, COMPACT-2 can be run in $O\left(\frac{N}{pB}+\log\frac{N}{B}\right)$ I/O rounds.

\subsubsection*{Problem $\mathcal R$ of Section~\ref{sec:prelim}}

Using COMPACT-2, PREFIX-SUMS-1, PREFIX-SUMS-2 and MERGE-3, we obtain the following lemma:

\begin{lemma}
\label{lem:aloc2}
On a CREW PEM with $p$ processors, the problem $\mathcal R$ can be solved in
	$O(\frac{N+kB}{pB} + \log k       + \log\log \frac{N}{B})$ I/O rounds,  if the input is sparse; and
	$O(\frac{N}{pB} + \log\frac{k}{B} + \log\log \frac{N}{B})$ I/O rounds,  if the input is compact. 
\end{lemma}

\proof
Please refer the solution to $\mathcal R$ presented in Section~\ref{sec:prelim}.
If the input array $A$ is compact, then keep arrays $E$, $D$ and $P$ all in compact form; the bound follows.
If $A$ is sparse, then invoke COMPACT-2 to get a compacted version of it; then proceed as above. 
\qed

Procedure Stratify can be used to design efficient algorithms for sorting and multiway-merging on
CREW PEM.

\subsection{Procedure Stratify on CREW PEM with $\sigma=B$ and $\lambda>B$}
\label{sec:stratpem}

Consider the procedure call ``$\stratify(N, k, X_0, \ldots, X_{k-1}, B, \lambda)$'', where $X_0, \ldots, X_{k-1}$ are sorted 
arrays of sizes $N_0, \ldots, N_{k-1}$, given contiguously one after another in that order in the shared external memory.
Without loss of generality, assume that each block of the input spans over at most three of the input arrays,
because, if a block spans $X_i, X_{i+1}, \ldots, X_{j-1}, X_j$, with $X_i$ and $X_j$ spanned partially, for $j> i+2$,
then $X_{i+1}, \ldots, X_{j-1}$ can
all be sorted together in place in one I/O by a single processor. Therefore, $k=O(N/B)$.

We can both partition the input arrays and 
	form the bristle arrays in $O(1)$ I/O rounds, using $N/B$ processors, once
	they are assigned one per block. 
(With $\lambda>B$, at most three bristles can be selected from every block; 
if a block spans $X_i, X_{i+1}$ and $X_{i+2}$, the last elements of all of them 
could be bristles and belong to it.)
By Lemma~\ref{lem:aloc2}, 
	with $p$ processors, this takes $O\left(\frac{N}{pB} + \log\log \frac{N}{B} + \log k\right)$ I/O rounds.
An invocation of prefix sums over $\langle \lceil \frac{N_0}{\lambda}\rceil, \ldots, \lceil\frac{N_{k-1}}{\lambda}\rceil\rangle$
	can be used to put the bristle arrays in sparse presentation, with the bristles distributed one per block.
Cross-ranking of the bristle arrays involves $\frac{k(k-1)}{2}$ merges, 
	each of size at most $s<\frac{2N}{\lambda}$, to	be performed in parallel. 
Each bristle participates in $k-1$ merges.
Use Lemma~\ref{lem:aloc} to assign $\Theta(\frac{p}{s})$ processors per bristle;
	$1/k$-th of those processors will participate in each of the merges of the bristle. 
Solve each merge instance independently using MERGE-2.
The time taken is $O(\mr{\frac{N}{\lambda}}{\frac{N}{\lambda}*\frac{p}{sk}})=O(\mr{\frac{N}{\lambda}}{\frac{p}{k}})=O(\frac{Nk}{p\lambda}+\log\log\frac{N}{\lambda})$ I/O rounds.

Steps~2 and 4 each requires every bristle to perform an addition of size $k$,
	of $\log n$ bit numbers, which are in sparse presentation.  
There are $s$ bristles. 
With $\Theta(\frac{p}{s})$ processors per bristle, 
	find the sums using PREFIX-SUMS-1 in $O(\frac{ks}{p}+\log k)$ I/O rounds.  

Each bristle has to perform $k$ searches in Step~3, each over a segment of
	size $\lambda$; that is, a total of $ks$ searches have to be performed 
	in parallel. 
At most $\Theta(\frac{p}{ks})$ processors are available per search.
Hence, using SEARCH-3, we can finish this step in $O\left(\frac{\log(\lambda/B)}{\log(p/ks)}\right)=
O\left(\frac{\log(\lambda/B)}{\log(p\lambda/Nk)}\right)$ I/O rounds.

Alternatively,
with $k$ processors, an array of size $\lambda$ can be searched in $O\left(\frac{\log(\lambda/B)}{\log k}\right)$ I/O rounds. With $p/ks$ processors, this search can be simulated in $O\left(\frac{k^2s}{p}.\frac{\log(\lambda/B)}{\log k}\right)$ I/O rounds.

In Step~5 we rebalance the bristles so that the number of elements associated with each surviving bristle is $\Theta(\sigma)$.
The \emph{value} of a bristle $L_j$ is 
$I(L_{j+1})-I(L_j)$,
the number of elements in its charge.
We classify the bristles as follows: (i)~\emph{small}: bristles with value less than $\sigma$, (ii)~\emph{large}: all remaining bristles.
If a \emph{small} bristle is flanked by \emph{large} bristles on both sides, then move its elements to the smaller of the two \emph{large} bristles. 
The receiving bristle would then have at most $k\lambda+\sigma$ elements in its charge. 
Partition $L$ into segments of size $\sigma$ and assign $p\sigma/s$ processors to each segment.
Within each segment we consider every maximal contiguous run of \emph{small} bristles and compute the prefix sums of their values.
Within these runs, we upgrade certain \emph{small} bristles to \emph{large} so that the number of elements between consecutive \emph{large} bristles becomes $\Theta(\sigma)$.
Whenever a prefix sum first reaches or exceeds a multiple of $3\sigma$, the corresponding bristle is labelled \emph{large}. 
Formally, for every pair of consecutive prefix sums $a,b$ satisfying
$a < 3\sigma r \leq b$
for some integer $r$, the bristle corresponding to $b$ is labelled \emph{large};
in this case, we say $r$ causes $b$ to be labelled \emph{large}.
Suppose integer $r$ causes a bristle to be labelled \emph{large}. 
Consider the first two prefix sums $\geq 3\sigma r$ that are labelled \emph{large}.
The smaller of them is in
$[3\sigma r,\;3\sigma r-1+\sigma-1]$, whereas
the prefix sum immediately preceding the larger of them  is in
$[3\sigma (r+1)-(\sigma-1),\;3\sigma (r+1)-1]$.
It follows that the value accumulated between consecutive \emph{large} bristles lies in the range
\(
[\sigma+3,\;3\sigma-1].
\)
Remove all \emph{small} bristles, and compact the remaining ones to form $L'$. With $\sigma=B$, the entire step requires $O(\frac{s}{p})$ I/O rounds.

The subtractions in Step~6 can be performed in $O(1)$ I/O rounds with $ks$ processors.
With $p$ processors, this would take
	$O(ks/p)$ I/O rounds.
Step~6 also requires every bristle to perform a prefix summation of size $k$,
	of $\log \lambda$ bit numbers. 
With a processor allocation similar to that in Step~2,
	the sums can be found in $O(\frac{ks}{p}+\log k)$ I/O rounds. 
Since the bristles are now merged into the array $L$, the allocation incurs
	only $O(1)$ time.

In Step~6, we also prepare the subarrays $S_{L_j}$.
Here we proceed as we did in the case of CREW
	PRAM; however some modifications are necessary to respect block boundaries. 
When some elements of a block are to be included in $S_{L_j}$, we would initially copy the entire block.
For bristle $L_j$, for $0\leq i\leq k-1$, 
	the $\delta_{i}(L_j)$ elements that are strictly between $L_j$ and $L_{j+1}$
	would span at most $\left\lceil\frac{\delta_{i}(L_j)}{B}\right\rceil+1$ blocks.
(For brevity, let $x_j$ denote $\sum_{i=0}^{k-1}\delta_{i}(L_j)$.)
Therefore, the number of elements copied into $S_{L_j}$ is at most $x_j+2kB$.
Over all the bristles, the number of elements copied is at most $N+2kBs =O\left(N\left(1+\frac{kB}{\lambda}\right)\right)$.
Divide the available processors into two equal lots. 
From one lot assign $\Theta(\frac{p}{s})$ processors to each bristle.
As in Step~1, distribute the other lot over the bristles so that $L_j$ gets $\Theta(\frac{x_j p}{N})$ processors.
Thus, $L_j$ gets a total of $\Theta(\frac{p(\lambda+x_j)}{N})$ processors.
The number of I/O rounds for the copying is $O\left(\max_{j}\frac{N}{pB}.\frac{(x_j+2kB)}{(\lambda+x_j)}\right)$.
These arrays have their unwanted elements marked and compacted out next. 
With the same processor distribution this would take  
$O\left(\max_{j}\left\{\frac{N}{pB}.\frac{(x_j+2kB)}{(\lambda+x_j)}  + \log\left(\frac{x_j+2kB}{B}\right)\right\} \right)$ I/O rounds.
Since $x_j<k\lambda$, Step~6 would take $O(\frac{N}{pB} + \frac{N}{p\lambda} + \log (k+\frac{k\lambda}{B}))$ I/O rounds. 

In Step~7, we copy the $S_{L_j}$'s into $X$ to form the $Y_j(X)$'s. 
Since each of them has a size of at least $\sigma=B$,
	the copying can be done in $O(N/pB + N/p\lambda)$ I/O rounds;
    the second term signifies the cost of handling the fractional blocks in parallel.

\begin{lemma}
\label{lem:7}
On a PEM that allows concurrent reads,
the procedure call `` $\stratify(N, k, X_0, \ldots, X_{k-1}, B, k^{2}B)$'' takes \(O\left(\frac{N}{pB}+\log\log \frac{N}{B} + \log k\right)\) I/O rounds with $p<N/B$ processors,
while `` $\stratify(N, k, X_0, \ldots, X_{k-1}, B, kB)$'' takes \(O\left(\log\log \frac{N}{B} + \log k\right)\) I/O rounds with $p\geq N/B$ processors,
\end{lemma}

\proof If $p\geq N/B$, then use first implementation of Step 3; otherwise, use the second one. \qed

\subsection{Sorting on CREW PEM}
\label{pem:segre}

Consider Procedure \segr\ described in Section~\ref{sec:anastratify}.
From Lemma~\ref{lem:7} it follows that an invocation \segr$(N,k,A,B,kB)$ of it runs in
\(O\left(\log\log \frac{N}{B} + \log k \right)\)
I/O rounds with $\frac{N\beta}{B}$ processors
	on a PEM that allows concurrent reads, where $\beta\geq 1$.

We now give the construction of a sorting scheme as before.
The basic building block of the scheme is MERGE-3 of Subsection~\ref{basicpem}; this 
	merges two arrays of a total size $N$ in 
	$O(\frac{N}{pB}+\log\log \frac{N}{B})$ time with $p$ processors on a CREW PEM; 
	let $a$ be the constant hidden by the $O$-notation here.
Guided by a complete binary tree on $N/B$ leaves, proceeding bottom up, and using
	this algorithm at each level, 
	the standard merge sort thus runs in $a\log \frac{N}{B}\log\log \frac{N}{B}$ time 
	on an $\frac{N}{B}$ processor CREW PEM.
This is the first algorithm in our sequence.
We denote this algorithm by \msort{0}{I}{N}. 

For $h>0$, \msort{h}{I}{N} is given below.
We use the notation introduced in the previous sections. 
For level \(j \ge 0\), define the $i$-th subarray of size $k_h^{j+2}$ in $I$ by
\( I^{(j)}_i(h) \;=\;
I\big[\, i k_h^{\,j+2} \ldots (i+1)k_h^{\,j+2}-1 \,\big],
\)
where \(0 \le j \le \log_{k_h}n - 2\) and \(0 \le i < n/k_h^{\,j+2}\).
Each subarray at level \(j\) is composed of \(k_h\) consecutive subarrays at level \(j-1\):
\(
I^{(j)}_i(h)
\;=\;
\bigcup_{t=ik_h}^{(i+1)k_h-1} I^{(j-1)}_t(h).
\)

\begin{figure}[h]
\label{msfig}
\vspace{0.1in}
\hrule
\vspace{0.1in}

\begin{tabbing}
aaaa \= aaaa \= aaaa \= aaaa \= aaaa \= aaaa \= \kill
{\bf Procedure \msort{h}{I}{N}} \\
\noindent
begin \\
     \> Let $n=N/B$; \\
     \> $k_h \leftarrow (\sqrt{\log n})^{f(h,n)}$, where $f(h,n)=1/\log^{(\lfloor h/2\rfloor+3)}n$; \\
     \> for $0\leq i< n/k_h^{2}$ in parallel do~~~~ 
		/* $I^{(0)}_{i}(h)=I[ik_h^{2},\ldots,(i+1)k_h^{2}-1]$ */ \\
     \>      \> \msort{h-1}{I^{(0)}_{i}(h)}{k_h^{2}}; \\
     \> for $j\leftarrow 1$ to $(\log n/\log k_h)-2$ do \\
     \> begin \\
     \>	     \> for $0\leq i< n/k_h^{j+2}$ in parallel do \\
     \>      \> begin \\
     \>      \>      \> \segr$(k_h^{j+2},k_h,I^{(j)}_{i}(h),B, k_hB)$; \\
     \>      \>      \> for $1\leq x\leq k_h^{j+1}$ in parallel do \\
     \>      \>      \>   \> \msort{h-1}{G_{x}(I^{(j)}_{i}(h))}{g_{x}(I^{(j)}_{i}(h))}; \\
     \>      \> end \\
     \> end \\
end 
\end{tabbing}

\vspace{0.1in}
\hrule
\vspace{0.1in}
\end{figure}

\noindent
Thus, with one processor per input block, for $n=N/B$,
	the procedure call ``\segr$(k_h^{j+2},k_h,I^{j}_{i},B,k_hB)$'' takes 
	$O(\log k + \log\log n)$ time on a CREW PEM.
Let $S_{h}(n)$ be the time taken by \msort{h}{I}{N} with 
	$N/B$ processors. 
Then, for $h>0$, 
    \[S_{h}(n) \leq  
		O(1) + S_{h-1}\left(k_h^{2}\right) +
		\left(\frac{\log n}{\log k_h}-2\right) 
		\left(O(\log k_h) +  O(\log\log n) + S_{h-1}\left(k_h^{2}\right)\right)\]
Therefore, the rest of the analysis can proceed as in the case of CREW PRAM.

Thus, we have the following theorem:

\begin{theorem}
\label{th:pemsort}
    Using one of the procedures \msort{2h}{I}{n}, for $h\geq 1$,  an array $I$ of size $N$ can be sorted 
	with $N/B$ processors, on a CREW PEM, in
	$O(\log \frac{N}{B})$ time. 
\end{theorem}

Arge et al.~\cite{Arge1} give bounds for sorting in the PEM model with \(p\) processors, including a lower bound of \(\Omega(\log (N/B))\) parallel I/Os. Aggarwal and Vitter~\cite{Aggvitt} show a lower bound of \(\Omega\!\left(\frac{N}{B}\log_{M/B}\frac{N}{B}\right)\) in the external memory model, implying \(\Omega\!\left(\frac{N}{pB}\log_{M/B}\frac{N}{B}\right)\) in PEM. Thus, the combined lower bound is
\[
\Omega\!\left(\frac{N}{pB}\log_{M/B}\frac{N}{B} + \log (N/B)\right).
\]
For \(p \geq n/\log m\), this is \(\Omega(\log n)\). Our algorithm matches this bound for \(p \geq n\). The upper bounds of Arge et al.~\cite{Arge1} are tight only for \(p < n/\log m\); hence, our algorithm is faster at higher parallelism.

\subsection{Multiway-merging on CREW PEM}

Given $k$ sorted arrays $X_0, \ldots, X_{k-1}$ of sizes $N_0, \ldots, N_{k-1}$, contiguously one after 
another in that order in the shared external memory, to merge them into a single sorted array with
$p$ processors proceed as follows:

Invoke the procedure call ``$\stratify(N, k, X_0, \ldots, X_{k-1}, B, k^{2}B)$''.
The output array $X$ has $s< 2N/k^2B$ (not necessarily sorted) 
	subarrays $Y_{j}(X)$ of size $|Y_{j}(X)| \in [B, (k^3+1)B]$,  for $0\leq j\leq s-1$, given one after another contiguously,
	so that for all $j$ and all $j'<j$, each element of $Y_j(X)$ is larger than every element of 
	$Y_{j'}(X)$. 
If subarrays $Y_{j}(X)$ is sorted in place, then $X$ would be sorted, as required.
Suppose $\Theta(\frac{p}{N}.|Y_{j}(X)|)$ processors are assigned to the $j$-th subarray.
The sorting can be completed in $\st{k^{3}B}{\frac{pk^{3}B}{N}}=O(\st{kB}{\frac{pkB}{N}})$ I/O rounds.

Therefore, we have the following lemma:

\begin{lemma}
Given $k$ sorted arrays $X_0, \ldots, X_{k-1}$ of sizes $N_0, \ldots, N_{k-1}$, contiguously one after 
another in that order in the shared external memory, the merge of them into a single sorted array can be computed 
in 	\[O\left(\frac{N}{pB}+\log\log \frac{N}{B} + \log k + \st{kB}{\frac{pkB}{N}} \right)\] I/O rounds 
	with $p$ processors on a PEM that allows concurrent reads.
\end{lemma}

\begin{cor}
If $p\leq N/B^2$ and $M=B^{O(1)}$, then by Corollary 4 of \cite{Arge1}, the above bound becomes
\[O\left(\frac{N}{pB}\log_{M/B}k + \log\log \frac{N}{B} + \log k \right)\] I/O rounds 
	with $p$ processors on a PEM that allows concurrent reads.
\end{cor}

\begin{cor}
If the sorting algorithm of Theorem~\ref{th:pemsort} is used, then the above bound becomes
\[O\left(\frac{N}{pB}\log k + \log\log \frac{N}{B} + \log k \right)\] I/O rounds 
	with $p$ processors on a PEM that allows concurrent reads.
\end{cor}

\section{Conclusion}
\label{sec:conclusion}

In this paper, we have presented efficient parallel algorithms for the sorting problem under three different computational models. Our main results are the following:

\begin{enumerate}
    \item an $O(\log n)$ time algorithm for sorting using $n$ processors on a CREW PRAM.
    
    \item an $O \left( \frac{\log n}{\log \alpha} \right)$ time algorithm for sorting using $n\alpha$ processors on the Parallel Comparison Model (PCM), where $2 \leq \alpha \leq n$.
    
    \item an algorithm for sorting in the Parallel External Memory (PEM) model with $p\geq \frac{N}{B}$ processors that requires $O \left( \log \frac{N}{B} \right)$ I/O rounds under the CREW constraint.
\end{enumerate}

Recursive versions of the first two algorithms, along with closing the gap between the lower and upper bounds in the PEM model, remain directions for future work.

\bibliographystyle{ACM-Reference-Format}
\bibliography{acm}

\end{document}